\documentclass[draft]{agujournal2019}

\usepackage{url} 
\usepackage{lineno}
\usepackage[inline]{trackchanges} 
\usepackage{soul}

\journalname{Geophysical Research Letters}

\begin{document}

\title{On the evaluation of solar wind's heating rates}

\authors{A. Zaslavsky \affil{1}}

\affiliation{1}{LESIA, Observatoire de Paris, Université PSL, CNRS, Sorbonne Université, Université de Paris, France}

\correspondingauthor{Arnaud Zaslavsky}{arnaud.zaslavsky@obspm.fr}



\begin{keypoints}
\item The use of least-square fittings of plasma data is shown to lead to unreliable evaluations of solar wind's heating rates.
\item The evolution of the adiabatic invariants from Helios proton core data shows clear evidence for perpendicular heating.
\item No statistically significant departure from adiabaticity is observed for protons in the direction parallel to the magnetic field.
\end{keypoints}

%
%

%
%


\begin{abstract}
Solar wind heating rates have often been calculated by fitting plasma and magnetic field data with a set of model functions. In this letter, we show that the rates obtained by such an approach strongly depend on the rather arbitrary choice one makes for these model functions. An alternative approach, consisting in monitoring the radial evolution of the adiabatic invariants, based on locally and consistently measured plasma and magnetic field parameters, is free from such a flaw. We apply this technique to a recently released Helios proton dataset, and confirm the existence of a clear perpendicular heating of solar wind's protons. On the other hand, no significant change in the parallel adiabatic invariant is visible in the data. We conclude that to date, and in the distance range of 0.3 to 1 AU, no clear observation of a deviation of solar wind's protons from parallel adiabaticity has ever been made.
\end{abstract}

\section*{Plain Language Summary}
The thermal expansion of the solar atmosphere into the interplanetary space produces the solar wind. This letter deals with the thermodynamics of the solar wind's expansion. An isolated gas is expected to get cooler as its expands. This is because its internal energy is consumed to power the expansion. But observations by space probes of the solar wind's temperature profile show that its decrease is less steep than expected. This means that the solar wind is heated by some external system -- which is suspected to be the interplanetary electromagnetic field. It is important to be able to properly quantify the amount of energy flowing into the solar wind along its expansion. This letter shows that a methodology used in several previous articles leads to unreliable estimations of the heating rates, because this methodology involves the use of arbitrary and possibly incompatible models for the magnetic field and plasma data. The application of a more robust methodology to a solar wind proton data supplied by the Helios spacecraft shows that if the solar wind is unambiguously heated in the direction perpendicular to the magnetic field, no convincing evidence of energy exchange exists in the direction parallel to it.

%
%

\section{Introduction}

The data provided by the Helios spacecraft allowed the first reconstruction of the density and temperature radial profiles of different populations of particles constituting the solar wind \cite{Marsch_etal_1982_p, Marsch_etal_1982_a}. Discrepancies between these profiles and the ones theoretically predicted for a spherically symmetric adiabatic expansion were observed, and naturally interpreted as due to energy flowing into the plasma along its expansion. The identification of the energy source and mechanisms of heating requires the heating rates to be carefully derived from the spacecraft data, an issue addressed by an abundant literature \cite<e.g.>{Freeman_1988, Cranmer_etal_2009, Hellinger_etal_2011_fast, Hellinger_etal_2013_slow, Stverak_etal_2015_elec, Perrone_etal_2019a, Perrone_etal_2019b}. This letter deals with the identification of a proper methodology to make such a derivation.

It is important to note that the spread in the plasma data collected at a given distance from the Sun is usually very important, reflecting the important space and time variability of solar wind's boundary conditions down in the corona. This spread implies large uncertainties on the calculated radial gradients, and therefore on the calculated heating rates. In order to overcome this problem, the procedure adopted in most previous works consisted in producing power law least square fitting of the density and the temperature radial profiles, and then to compute the gradients of these quantities from the obtained power law indices -- therefore neglecting the spread and more or less implicitly assuming the plasma to expand according to a polyropic law \cite{Freeman_1988, Totten_etal_1995}. The obtained power-law exponents are finally combined with a magnetic field model in order to derive the anisotropic heating rates that are relevant for a magnetized plasma expansion.

An alternative approach, pioneered by \citeA{Marsch_eqstate_1983}, consists in the calculation of the heating rates from the radial evolution of the CGL adiabatic invariants \cite{CGL_1956}. This approach yields result incompatible with the one previously mentionned: while the ``polytropic method'', implemented by \citeA{Hellinger_etal_2011_fast, Hellinger_etal_2013_slow} for solar wind protons with a Parker spiral magnetic field model, concluded in the existence of a proton parallel cooling close to the Sun, that is progressively changed into a heating farther away from it, \citeA{Marsch_eqstate_1983}, on a similar dataset, observed the proton evolution to be essentially adiabatic in the parallel direction. In this letter, we show that the method employed by \citeA{Hellinger_etal_2011_fast, Hellinger_etal_2013_slow} is flawed by the strong dependency of the conclusions it provides on the choice made for the magnetic field model. On the other hand, the CGL method of \citeA{Marsch_eqstate_1983} only involves locally and consistently measured magnetic field and plasma parameters. Therefore, the conclusions it provides do not dependent on preconceptions about fitting functions used to model the data -- and we have to decide the inconsistency appearing in the literature in favor of a proton parallel adiabaticity, as first observed by \citeA{Marsch_eqstate_1983}.

In the section 2 of this letter, we remind some theoretical background on the energetics of the expansion of a magnetized plasma, necessary to the derivation of plasma heating rates from the data. In section 3, we apply the methodology of \citeA{Hellinger_etal_2011_fast, Hellinger_etal_2013_slow} to a Helios proton dataset, and show why the results of such a methodology cannot be relied on. In section 4 we monitor the radial evolution of the adiabatic invariants computed from the same dataset, and conclude that a proton heating is clearly occurring in the direction perpendicular to the magnetic field, but that the expansion is consistent with adiabaticity in the parallel direction. Section 5 summarizes our results.

\section{Theoretical background}

Under the assumption of gyrotropy of a particle velocity distribution function around the magnetic field $\textbf{B} = B \textbf{b}$, its second order centered moment (i.e. its pressure tensor) $\textbf{p}$ is diagonal and can be written as $\textbf{p} = p_\parallel \textbf{b}\textbf{b} + p_\perp (\textbf{I} - \textbf{b}\textbf{b})$. The evolution of the pressure components $p_\parallel$ and $p_\perp$ is given by -- see e.g. \cite{Chust_Belmont_2006}:
\begin{equation}
\frac{dp_\parallel}{dt}  + p_\parallel \nabla \cdot \textbf{u} + 2 p_\parallel \nabla_\parallel \cdot \textbf{u} = Q_\parallel - \nabla \cdot (q_\parallel \textbf{b}) + 2 q_\perp \nabla\cdot \textbf{b} \equiv Q'_\parallel
\label{eq_p_para}
\end{equation}
\begin{equation}
\frac{dp_\perp}{dt} + 2 p_\perp \nabla \cdot \textbf{u} - p_\perp \nabla_\parallel \cdot \textbf{u} = Q_\perp - \nabla \cdot (q_\perp \textbf{b}) - q_\perp \nabla\cdot \textbf{b} \equiv Q'_\perp
\label{eq_p_perp}
\end{equation}
where $\textbf{u}$ is the population's mean velocity, $q_\parallel$ and $q_\perp$ parallel and perpendicular components of its heat flux density vector, and $\nabla_\parallel = \textbf{b} (\textbf{b} \cdot \nabla)$ is the gradient along $\textbf{b}$. $Q_\parallel$ and $Q_\perp$ account for energy transfer (per unit volume and time) from external systems (for instance by collision with other populations of particles or by interaction with the electromagnetic field), or from the other direction of the same system (for instance by collisions internal to this population). $Q'_\parallel$ and $Q'_\perp$ are introduced as the right hand terms of these equations. These are the parameters that we are looking forward to derive from the data. The proper external heating rates can then be obtained from the measurement of the divergence of the heat flux. If this divergence is negligible -- which is typically the case for protons in the solar wind, see  \citeA<e.g.>{Hellinger_etal_2011_fast, Hellinger_etal_2013_slow} -- then $Q'$ is essentially the external heating rate. 

Under the supplementary assumption that the magnetic and velocity fields are linked through ideal Ohm's law (i.e. the solar wind conductivity can be considered as infinite at our scales of interest), one may express $\nabla_\parallel \cdot \textbf{u} = \nabla \cdot \textbf{u} + B^{-1}dB/dt$. Now using the continuity equation $dn/dt + n\nabla \cdot \textbf{u} = 0$ to eliminate the divergence of the velocity field from the equations, one may recast eqs.(\ref{eq_p_para})-(\ref{eq_p_perp}) as follows:
\begin{equation}
\frac{1}{p_\parallel} \frac{dp_\parallel}{dt} - \frac{3}{n} \frac{dn}{dt} +  \frac{2}{B} \frac{dB}{dt}  = \frac{d \ln C_\parallel}{dt} = \frac{Q'_\parallel}{p_\parallel}
\label{CGL_para}
\end{equation}
\begin{equation}
\frac{1}{p_\perp} \frac{dp_\perp}{dt} - \frac{1}{n} \frac{dn}{dt} -  \frac{1}{B} \frac{dB}{dt}  =  \frac{d \ln C_\perp}{dt}= \frac{Q'_\perp}{p_\perp}
\label{CGL_perp}
\end{equation}
were we introduced the notations $C_\parallel = p_\parallel B^2 /n^3$ and $C_\perp = p_\perp / nB$. The system's behavior is double-adiabatic if it exchanges no heat with external systems in either direction, and if the divergence of the heat flow is zero, i.e. when $Q'_\parallel = Q'_\perp = 0$. In this case one recognizes the famous equations from \cite{CGL_1956} describing the conservation, along the plasma expansion, of the quantities $C_\parallel$ and $C_\perp$. In this paper we shall say that the system shows parallel or perpendicular adiabaticity when one of these invariants is conserved, that is, if $Q'_\parallel$ or $Q'_\perp$ is equal to zero. Strictly speaking, this does only imply that there is no net energy transfer into the parallel or perpendicular subsystem, and does not exclude scenarios in which one of these subsystems would receive an external heating, which would be at every point compensated by an internal transfer of the same amount of energy into the other subsystem (i.e., from a direction to the other). Eqs.(\ref{CGL_para})-(\ref{CGL_perp}) relate the local heating rates to the variation of the adiabatic invariants. Since the total time derivatives in these equations cannot be evaluated (for this it would be necessary for the measurement device to follow the same plasma fluid element along its trajectory), the heating rates have to be calculated under the assumptions that the system is stationary and dependent only on the $r$ coordinate (invariant by rotations centered on the Sun), so that $d/dt \equiv u_r d/dr$. In this case the time derivatives are equivalent to radial gradients, which can be measured by a spacecraft with an orbit elliptical enough to scan a wide range of distances to the Sun.

In order to evaluate the heating rates, two ways, in apparence equivalent, now appear: the first is to evaluate the derivatives appearing in the left-hand side of eqs.(\ref{CGL_para})-(\ref{CGL_perp}) by approximating the parameters $p_\parallel$, $p_\perp$, $n$ and $B$ by some chosen functions of $r$, adapted to spacecraft data by a least-square fitting procedure. The second approach is to compute the adiabatic invariants $C_\parallel$ and $C_\perp$ from locally measured parameters, and then to derive the heating rates from the radial variation of these quantities. We shall in the following apply both of these methods to the same dataset. For this purpose, we chose a publicly available protons dataset measured by Helios \cite{Stansby_dataset_2017}. These data were obtained through a double Maxwellian fitting of the protons velocity distribution function core component, with a methodology detailed by \citeA{Stansby_SP_2018}. Importantly, it must be noted that, since this fitting only concerns the thermal core of the distribution function, no effect related to the radial evolution of the proton beam will appear in our treatment, so that an observed heating or cooling may not be interpreted as being due to a change, with radial distance, of the proton beam drift velocity with respect to the thermal core -- an effect that could play a role if we considered the evolution of the total equivalent temperature of the core and beam system.

\section{Evaluation of the heating rates from fitted plasma parameters and a magnetic field model}

\begin{figure}
\includegraphics[scale = 0.7]{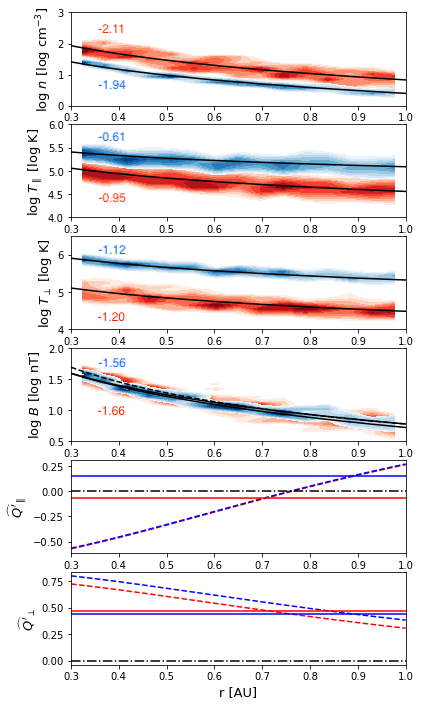}
\caption{Evolution of the proton parameters measured by Helios as a function of the distance to the Sun for slow (red contours) and fast (blue contours) winds. From top to bottom, the three first panels show the decimal logarithm of the plasma density, parallel and perpendicular temperature as a function of $r$. Over imposed are the power law best fits of the data, with their associated power-exponents. The fourth panel shows the magnetic field as a function of $r$. Over imposed is the power law best fits (solid line) for slow and fast winds, as well as a best fit for the Parker spiral model discussed in the text (dashed line). The last two panel shows the unreliable estimates of $\widehat{Q'_\parallel}$ and $\widehat{Q'_\perp}$ calculated from the plasma power exponents and different magnetic field models (solid lines: power-law model, dashed lines: Parker spiral model). The dot and dashed black line indicates $Q'=0$.}
\label{fig_evol_plasma_r}
\end{figure}

Let us start by introducing the local power exponent for the function $f(r)$ (which, in the following, will be either the density $n$, the temperatures $T_\parallel$, $T_\perp$ or the magnetic field modulus $B$) as $\alpha_f(r) = -(r/f)(df/dr)$. We also introduce the normalized heating rates $\widehat{Q'_\parallel} \equiv Q'_\parallel/(p_\parallel u_r/r)$ and $\widehat{Q'_\perp} \equiv Q'_\perp/(p_\perp u_r/r)$. Using these notations, eqs.(\ref{CGL_para})-(\ref{CGL_perp}) can be conveniently rewritten as
\begin{equation}
\widehat{Q'_\parallel} = - \alpha_{T_\parallel} + 2 \alpha_{n} -2 \alpha_B 
\label{Q_polytropic_para_exponents}
\end{equation}
\begin{equation}
\widehat{Q'_\perp} =  - \alpha_{T_\perp} + \alpha_B 
\label{Q_polytropic_perp_exponents}
\end{equation}
where the parallel and perpendicular temperature are defined with respect to the pressure tensor components according to $p_\parallel = nkT_\parallel$ and $p_\perp = nkT_\perp$. These equations show how the heating rates are related to the local power-exponents of the plasma parameters. From this, a seemingly straightforward approach to the calculation of the heating rates, employed by \citeA<e.g.>{Hellinger_etal_2011_fast, Hellinger_etal_2013_slow, Stverak_etal_2015_elec}, consists in deriving power exponents by least-square fitting the measured plasma parameters with functions of the form $f(r) \propto r^{-\alpha_f}$, $\alpha_f$ being a constant. Fig.\ref{fig_evol_plasma_r} shows the result of such an analysis performed on our dataset. To obtain this figure, the data collected by both Helios probes from 1974 to 1984 were first separated into fast wind (between 600 and 800 km.s$^{-1}$) and slow wind (between 290 and 380 km.s$^{-1}$). These velocity widths were chosen in order for both data sets to contain roughly the same number of samples ($\sim 120000$). The data were then binned in 15 distance intervals equally spaced between 0.3 and 1 AU, and histograms of the logarithm of the different parameters were computed in each of these intervals, using 50 regularly spaced bins. The histograms were then normalized in each distance bin, and their contours (per $10\%$ steps, starting from $30\%$) were plotted in red and blue for the slow and fast wind sets, respectively.

In the following, in order to keep the discussion reasonably short, we focus on the evaluation of the parallel heating rate in the fast wind sample. The same reasoning can, of course, be carried out for the perpendicular direction and the slow wind sample, we shall come back to it at the end of this section. The best-fit power exponents are $\alpha_{T_\parallel} = 0.61$, $\alpha_{n} = 1.94$ and $\alpha_B = 1.56$. Using these numbers and eq.(\ref{Q_polytropic_para_exponents}), we obtain a normalized parallel heating rate independent of $r$, with a value $\widehat{Q'_\parallel} = 0.15$: according to this first analysis, one should conclude that the protons undergo a parallel heating on the whole distance range scanned by Helios. As an aside, one may note that the value obtained is quite small, and that it is conceivable that these measurements are consistent with parallel adiabaticity.

The power exponents presented here are reasonably close to the ones derived by \citeA{Hellinger_etal_2011_fast} (i.e. $\alpha_{T_\parallel} = 0.54$, $\alpha_{n} = 1.8$ and $\alpha_B = 1.6$) on a fast wind dataset where the parallel temperature was defined as the total temperature, therefore including the drift energy of the proton beam as part of the internal energy. Using again eq.(\ref{Q_polytropic_para_exponents}), one can derive from the exponents of  \citeA{Hellinger_etal_2011_fast} a value $\widehat{Q'_\parallel} = -0.14$, that is, a constant normalized rate but this time negative. However, the authors of this paper did not conclude in a parallel cooling of the protons from $0.3$ to $1$ AU, but instead in a cooling close to the Sun and a heating farther away from it. The reason for this is that they did not use the constant power law exponent $\alpha_B = 1.6$ derived from the data to calculate the heating rates. Instead, they derived the value of $\alpha_B$ from a Parker spiral model. The modulus of the magnetic field is in this case given by \cite{Parker_IDP_1963}
\begin{equation}
B(r) \propto \frac{1}{r^2} \left( 1 +   \frac{r^2 \tan^2\psi(r_0)}{r_0^2}   \right)^{1/2} 
\label{B_Parker}
\end{equation}
where $\psi(r_0)$ is angle between the tangential and radial component of the field at the distance $r_0$. In their work, \citeA{Hellinger_etal_2011_fast} used $\psi(r_0 = 1\textrm{ AU}) = 45^\circ$, and we shall use the same value. For such a spiral model, the magnetic field power exponent will not be a constant on the whole distance interval, but vary from $\alpha_B \simeq 2$ for small values of $r$, to $\alpha_B \simeq 1$ for large values of $r$ (small and large here being defined with respect to $r_0 = 1$ AU). Keeping the values of $\alpha_{T_\parallel}$ and $\alpha_{n}$ from  \citeA{Hellinger_etal_2011_fast}, one now obtains a normalized heating rate $\widehat{Q'_\parallel}$ varying with distance, between the asymptotic values $\widehat{Q'_\parallel} \sim -0.94$ for small values of $r$ and $\widehat{Q'_\parallel} \sim 1.06$ for large values of $r$. Therefore, the decision to use the Parker spiral model rather than the power-law fitting of $B$ lead the authors of  \cite{Hellinger_etal_2011_fast} to conclude that a parallel cooling of the protons occurs in the vicinity of the Sun, and that a parallel heating occurs far away from it.

Of course, if we make the same decision and apply this Parker spiral model to our dataset, we shall also obtain a heating rate $\widehat{Q'_\parallel}$ varying with distance, in our case between asymptotic values $-0.73$ and $1.27$, for small and large $r$ respectively. The actual variation of $\widehat{Q'_\parallel}$ is shown by the black curve on the fifth panel of Fig.\ref{fig_evol_plasma_r}, and one can see that we recover a result similar to the one of \citeA{Hellinger_etal_2011_fast}, and are lead to conclude in the occurrence of a parallel cooling close to the Sun and heating farther away from it. The lower panel of Fig.\ref{fig_evol_plasma_r} shows that the same effect, of course, affects the perpendicular heating rates, the values of which strongly depend on the magnetic model used. However, the relatively high values of $\alpha_{T_\perp}$ make the heating rates to be positive, whatever the model used, on the whole $0.3-1$ AU interval. The model dependency of the result is then less striking in this direction.

The previous considerations show that the cooling close to the Sun and heating far away from it is not a feature characteristic of a particular dataset -- in particular, this effect is not linked, here, to the deceleration of the proton beam close to the Sun, as was envisaged by \citeA{Hellinger_etal_2011_fast}, since it is also seen in a dataset where the proton core only is considered. It is instead a feature of a particular choice of a magnetic field Parker spiral model. This feature completely disappears when the magnetic field model is changed to a power-law one, and the choice of yet another magnetic field model would lead to yet another behavior for the radial evolution of the heating rate.

From this clearly appears the intrinsic limitation of the use of eqs.(\ref{Q_polytropic_para_exponents})-(\ref{Q_polytropic_perp_exponents}) to evaluate the plasma heating rates: the values obtained, and the conclusion reached in terms of heating or cooling, depend drastically on the choice of the model used for $B$ -- and not much on the dataset used. Moreover, the fourth panel of fig.\ref{fig_evol_plasma_r} shows that both models (Parker spiral and power law model) are essentially consistent with the data points: this methodology provides us with no clear way to decide what is the proper magnetic field model to use, and, therefore, what proper conclusions on the plasma heating or cooling should be drawn. Pragmatically, one may only conclude that the methodology presented in this section does not provide us with a reliable way to estimate the solar wind plasma heating rates.

\section{Monitoring the radial evolution of the adiabatic invariants}

As we just exposed it, the fundamental reason of the unreliability of eqs.(\ref{Q_polytropic_para_exponents})-(\ref{Q_polytropic_perp_exponents}) to evaluate the heating rates is the strong dependency of the results obtained on the choice of a specific set of functions (e.g. power law and parker spiral, but the same criticism would apply to any different choice of functions) to model the plasma and magnetic field data. But we saw from eqs.(\ref{CGL_para})-(\ref{CGL_perp}) that there is no necessity, in order to determine the heating rates, to introduce any models for the variation of the plasma parameters with the distance to the Sun. We can instead compute the adiabatic invariants $C_\parallel$ and $C_\perp$ from the locally measured plasma and magnetic field parameters, and then monitor the evolution of these quantities with radial distance. Beyond the crucial point of avoiding the use of models for the plasma and magnetic field data, this approach also presents the advantage of preserving all the information on the correlations that may exist between the locally measured quantities $n$, $T_\parallel$, $T_\perp$ and $B$. An obvious, but important, example of such a correlation is the one that stems from the very definition of the parallel and perpendicular directions: it is clear that the use of eqs.(\ref{eq_p_para})-(\ref{eq_p_perp}) can be justified only if the magnetic field B is the one with respect to which the parallel and perpendicular components of the pressure tensor are defined. This is clearly ensured in the computation of the adiabatic invariants from the locally measured $B$ and $p_\parallel$ and $p_\perp$, which are all consistently defined. On the other hand the use $p_\parallel$ and $p_\perp$ data defined with respect to the local field is  \emph{a priori} inconsistent with the use for B of an averaged, e.g. Parker spiral, model.

\begin{figure}
\includegraphics[scale = 0.6]{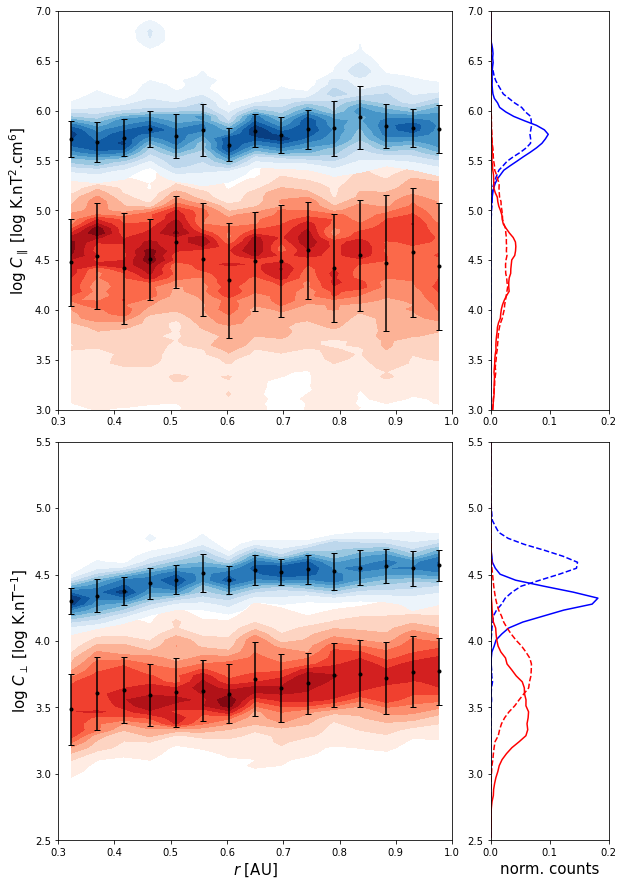}
\caption{Evolution of the logarithm of the adiabatic invariants $C_\parallel$ and $C_\perp$ as a function of the distance to the Sun for slow ($290<v<380$ km.s$^{-1}$, red contours) and fast ($600<v<800$ km.s$^{-1}$, blue contours) winds. Error bars overplotted show the mean and standard deviation of $\log C_\parallel$ and $\log C_\perp$ in 15 distance bins. The right panels show these distributions in the closest (solid line) and the farthest (dashed line) bins from the Sun.}
\label{Fig_C_cgl}
\end{figure}

Fig.{\ref{Fig_C_cgl}} shows the radial evolution of the parallel and perpendicular adiabatic invariants obtained from the dataset previously introduced. The presentation of the data is made according to the same methodology and color code as for Fig.\ref{fig_evol_plasma_r}. Additionally, the mean and standard deviations were computed in each distance bin from a gaussian best fit of the data (in order to provide values robust to the presence of outliers) and over-plotted on the contours as error bars.

The top panels of Fig.{\ref{Fig_C_cgl}} show the histograms of the parallel invariant $\log C_\parallel$, which seems reasonably constant along the expansion. Histograms of $\log C_\parallel$ in the closest and farthest distance bins from the Sun, as illustrated on the right panel of the figure, are very comparable. And if their mean values are not precisely the same, of course, difference between them are very small compared to their standard deviation. At the opposite, the bottom panels show a clear increase of $\log C_\perp$ with increasing radial distance, with differences in the mean values of the order of a standard deviation between 0.3 and 1 AU. The trend is a bit clearer in the fast than in the slow wind, not because the increase in $C_\perp$ is larger in average (it is about the same), but because the spread of the data is quite larger in the slow than in the fast wind.

Therefore we conclude from this analysis that no departure from adiabaticity in the parallel direction is visible in this proton dataset. This conclusion is consistent with the one of \citeA{Marsch_eqstate_1983}, that was also based on the adiabatic invariant monitoring, although their study was performed on a dataset in which the temperature is defined as the total temperature of the core and beam system. Taking or not the proton beam into account therefore does not seem to noticeably impact the conclusion that the proton expansion is close to be adiabatic in the parallel direction. On the other hand these results contradict those of \cite{Hellinger_etal_2011_fast, Hellinger_etal_2013_slow} -- which, as discussed in the previous section, appear as a methodological artifact.

We also conclude that a perpendicular heating of the protons can clearly be observed in the solar wind. The energy transfer rate associated with this heating can be evaluated from the slope measured in the data. The average slope on the whole distance interval is $d \ln C_\perp/dr \sim 1$ AU$^{-1}$ in the slow as in the fast wind sets, from which one obtains $Q'_\perp \sim p_\perp u_r / 1 \textrm{AU}$. One can maybe more conveniently express this heating rate per proton and per unit distance, and get a value $\sim kT_\perp / 1 \textrm{AU}$, which means that, whether in the slow or fast solar wind, and as a rough order of magnitude, a thermal proton gains its own perpendicular thermal energy while travelling an astronomical unit. Since the perpendicular proton heat flux divergence is several order of magnitudes below this value \cite{Hellinger_etal_2011_fast, Hellinger_etal_2013_slow} one can safely interpret this value as due to external heating. Wave-particle interactions or reconnection events both constitute credible candidates to produce this kind of anisotropic energy transfer.

Looking a bit more into the details of the variations, one may notice that, if the slope of $\ln C_\perp$ seems roughly constant in the slow wind set, in the fast wind set this variation seems to be occurring mostly between 0.3 and 0.6 AU, with a slope $\sim 2$ AU$^{-1}$, while the slope beyond 0.6 AU is in average zero. This result may lead us to think that the heating of the fast wind mostly takes place -- or at least is relatively stronger -- close to the Sun than far away from it. 

Let us finally note that it is probably dubious to try to go much further than these orders of magnitude estimations of the proton heating rates. This is due to the large dispersion of the observed data in each distance bin, which makes a precise estimation of the local slope in $\ln C_\perp$ rather hazardous. This dispersion is, probably to a large extent, linked to the methodology consisting in using all the data gathered by the Helios probes on a large time interval, and therefore mixing up solar wind streams with different origins in the same dataset. This is particularly true for the slow wind, where the distributions of the adiabatic invariants can be quite variable from a distance bin to another. This dispersion could probably be reduced by using spacecraft conjunctions to observe the same plasma parcel at different stages of its radial evolution \cite<e.g.>{Schwartz_Marsch_1983} but even in this case, the radial gradient in $\ln C_\perp$ will only be calculable on a length scale of the distance between the two spacecraft in conjunction, which will typically be of the order of the astronomical unit. Since spacecraft conjunctions are quite rare events, another way to go is to assume steadiness of coronal hole fast wind streams and identify crossings of the same stream at different distances from the Sun. Such a methodology was recently set up by \citeA{Perrone_etal_2019a} using Helios proton core data, and concluded, in agreement with \citeA{Marsch_eqstate_1983} and with the present study, that the perpendicular adiabatic invariant conservation was violated, but that no clear conclusion on the variation of the parallel invariant could be reached.

\section{Conclusion}

In this letter, we pointed out that a procedure for evaluating solar wind's plasma heating rates, used in several previous papers, based on fitting the plasma and magnetic field data to a set of model functions -- typically power laws and Parker spiral models -- leads to unreliable results. This is because the results it provides strongly depend on a rather arbitrary choice of model functions to fit the data. In particular, we show that the proton parallel cooling/heating pattern observed by \citeA{Hellinger_etal_2011_fast, Hellinger_etal_2013_slow} is a direct consequence of the use of a Parker spiral model for the magnetic field, and that the effect disappears if a power-law model is instead applied. We also show that this Parker spiral model, when used to derive heating rates from a dataset including only the Maxwellian core of the proton population, gives rise to the same parallel cooling/heating pattern. This shows that this pattern is an artifact from the method used, and not a consequence of the slowing down of proton beams close to the Sun. That said, effects related to the proton beam dynamics may still exist -- see for instance the study of Ulysses data in the fast wind between 1.5 an 4 AU by \citeA{Matteini_etal_2013} -- but they are still to be identified in Helios data. This clearly motivates future studies involving separate fittings of the core and the proton beams, in order to track separately the evolution of these systems.

On the other hand, the adiabatic invariant approach to the evaluation of plasma heating rates involves only operations on locally measured parameters, and does not involve any modeling of the data. It is therefore free from the bias identified in the previous method, and the results it provides should be taken as more reliable. The application of this method shows an increase of the perpendicular adiabatic invariant between 0.3 and 1 AU, confirming the existence of a local perpendicular proton heating in the solar wind. The order of magnitude of the associated heating rate (per unit volume) is $Q_\perp \sim p_\perp u_r / 1 \textrm{AU}$ in the slow as in the fast wind. No significant change in the adiabatic invariant can be observed in the parallel direction, and one cannot exclude, on the basis of our dataset, that the expansion occurs adiabatically along the magnetic field.

Since Helios proton core data are compatible with parallel adiabaticity, and since the studies including proton beams were either lead with a problematic methodology \cite{Hellinger_etal_2011_fast, Hellinger_etal_2013_slow}, either concluded in an almost conserved parallel invariant \cite{Marsch_eqstate_1983}, we have to conclude that, to date, and in the distance range of 0.3 to 1 AU, no clear observation of a deviation of solar wind's protons from parallel adiabaticity has ever been made. The investigation of Parker Solar Probe's data may either confirm or deny this trend in a closer neighborhood to the Sun. We can only insist, of course, that such an investigation be done by tracking the radial evolution of the CGL adiabatic invariants.

\section*{Open Research}
The Helios proton data used in this paper is available through \citeA{Stansby_dataset_2017} and \citeA{Stansby_SP_2018}. 

\acknowledgments
The author would like to thank the two referees, whose reviews helped to improve considerably the final version of this article.

\bibliography{heating}

%
%
%
%
%

\end{document}